\documentclass[twocolumn,showpacs,preprintnumbers,amsmath,amssymb,superscriptaddress]{revtex4}
\usepackage{epsfig}
\usepackage{graphicx}
\begin{document}
\title{Mirror symmetry breaking as a problem in dynamic critical phenomena}
\author{David Hochberg}
\email{hochbergd@inta.es} \homepage[]{http://www.cab.inta.es}
\affiliation{Centro de Astrobiolog\'{\i}a (CSIC-INTA), Ctra. Ajalvir
Km. 4, 28850 Torrej\'{o}n de Ardoz, Madrid, Spain}
\author{Mar\'{i}a Paz Zorzano}
\email{zorzanomm@inta.es} \affiliation{Centro de Astrobiolog\'{\i}a
(CSIC-INTA), Ctra. Ajalvir Km. 4, 28850 Torrej\'{o}n de Ardoz,
Madrid, Spain}

\date{\today}
\begin{abstract}
The critical properties of the Frank model of spontaneous chiral
synthesis are discussed by applying results from the field theoretic
renormalization group (RG).  The long time and long wavelength
features of this microscopic reaction scheme belong to the same
universality class as multi-colored directed percolation processes.
Thus, the following RG fixed points (FP) govern the critical
dynamics of the Frank model for $d<4$: one unstable FP that
corresponds to complete decoupling between the two enantiomers, a
saddle-point that corresponds to symmetric interspecies coupling,
and two stable FP's that individually correspond to unidirectional
couplings between the two chiral molecules. These latter two FPs are
associated with the breakdown of mirror or chiral symmetry. In this
simplified model of molecular synthesis, homochirality is a natural
consequence of the intrinsic reaction noise in the critical regime,
which corresponds to extremely dilute chemical systems.
\end{abstract}

\pacs{05.70.Jk, 64.60.Ak, 05.10.Gg}

\maketitle

\section{\label{sec:intro} Introduction}

Mirror symmetry is broken in the bioorganic world and life as we
know it is invariably linked to biological homochirality. An
outstanding problem associated with the origin of life is to explain
chiral symmetry breaking in nature, why for example, it came to be
that the nucleotide links of RNA and DNA incorporate exclusively
righthanded sugars while the enzymes involve only the lefthanded
amino acids.  A recent survey of hypotheses concerning this
phenomenon, experimental realizations and additional pertinent
bibliography can be found in the references
\cite{AG,Avalos2,Kondepudi,Blackmond,WLL}.

The essential key ingredients of theoretical models of
mirror-symmetry breaking processes in chemistry \cite{Blackmond}
include reactions in which the products serve as catalysts to
produce more of themselves while inhibiting the production of their
chiral or mirror-image counterparts. In chemistry,
\textit{enantiomers} are molecules that are nonsuperimposable
complete mirror images of each other. Frank's original model
\cite{Frank}, and a variant of which we study in this paper,
involves autocatalysis of the two enantiomers, denoted here as L and
D \cite{footnote1}, and mutual inhibition or antagonistic effects
between the two chiral species. More recently, Sandars introduced a
model in which the detailed polymerization process and enantiomeric
cross-inhibition are taken into account, its basic features are
explored numerically, but without including spatial extent, chiral
bias or noise \cite{Sandars}. Brandenburg and coworkers have
analyzed further properties of Sandars' model and have proposed a
truncated version including chiral bias \cite{BAHN}, and have
studied this reduction with spatial extent and coupling to a
turbulent advection velocity \cite{BM}. Gleiser and Thorarinson
analyze the reduced Sandars' model with spatial extent and coupling
to an external white noise \cite{GT} and in \cite{Gleiser}, Gleiser
considers the reduced chiral biased model with external noise.
Despite the simplicity of the Frank model, ignoring as it does the
polymerization process, it continues to serve as a kind of ``Ising
model" for chiral symmetry breaking, and the purpose of this paper
is to better understand its critical properties by exploiting the
model's relation to directed percolation phenomenology.

The specific reaction scheme we will study in this paper is given as
follows. The $k_i$ denote the reaction rate constants and we take
the achiral substance A as a uniform constant background.

\noindent
Autocatalytic production:
\begin{equation}\label{autoLD2}
\textrm{L} + \textrm{A} \stackrel{k_1}{\rightleftharpoons
\atop{\small k_3}}\textrm{L} + \textrm{L}, \qquad \textrm{D} +
\textrm{A} \stackrel{k_1} {\rightleftharpoons \atop{\small k_3}}
\textrm{D} + \textrm{D}.
\end{equation}
Dimerization and additional mutual inhibition in second order
reactions:
\begin{equation}\label{mutualgeneral}
\textrm{L} + \textrm{D} \stackrel{k_2}{\longrightarrow} P, \qquad
\textrm{L} + \textrm{D} \stackrel{k_4}{\longrightarrow} \textrm{L} +
\textrm{A}, \qquad \textrm{L} + \textrm{D}
\stackrel{k_5}{\longrightarrow} \textrm{D} + \textrm{A}.
\end{equation}
Spontaneous decay or recycling back to the achiral substrate:
\begin{equation}\label{decay}
\textrm{L} \stackrel{k_6}{\longrightarrow} \textrm{A}, \qquad
\textrm{D} \stackrel{k_6}{\longrightarrow} \textrm{A}.
\end{equation}

The above scheme differs from the original Frank model \cite{Frank}
in the open-flow reactor nature of the process and the fact that the
reaction Eq.(\ref{autoLD2}) is allowed to be reversible ($k_3 \geq
0$).  We assume that each enantiomer diffuses with the same
diffusion constant $D_0$ and incorporate this feature into the
master equation description of this process. We also account for two
inhibitory or mutually antagonistic reactions, with associated rates
$k_4$ and $k_5$, in addition to Frank's dimerization step, $k_2$.
This scheme is a partial hybrid between the Frank model and the
Avetisov and Goldanskii (AG) reaction (see e.g., Eq.(13) of
\cite{AG}). Whereas Frank's original model gives rise to pure
homoquiral states in which only one enantiomer is present, the
complete AG model leads to chiral symmetric broken final states
where both enantiomers are present in unequal proportions. Mirror
symmetry is broken in the AG model, but the breaking is not
absolute.

Both the above and Frank's 1953 scheme yield the same
field-theoretic structure for the \textit{effective action}, and
more importantly, therefore belong to the same \textit{universality
class}. In our analysis, we allow for $k_4 \neq k_5$, as this leads
to a rich fixed point structure in the critical regime of the model.
This choice implicitly accounts for the influence of an external
chiral field or bias. Of course, a chiral symmetric action and
Langevin equations result from the ``natural choice" $k_4 = k_5$.
The properties of the chirally unbiased model can be studied as a
special case of the above scheme.

Chiral or mirror symmetry breaking is an example of a nonequilibrium
phase transition which is attained when the control variable $(k_1A
- k_6) \approx 0$ becomes small. A continuous, second order,
transition is then induced from a fully active state, characterized
by the simultaneous presence of both competing enantiomers
accompanied by fluctuations in each chirality, to an inactive or
absorbing state, in which only one enantiomer survives. This control
parameter can be made small (or large) by simply adjusting the
concentration of the achiral molecule $A$. This limit implies that
the net amount of total chiral matter is vanishingly small, i.e.,
the chemical system is extremely dilute at criticality. This is
because the autocatalytic amplification of the enantiomers is
delicately balanced by their spontaneous decay. The purpose of this
paper is to understand this specific critical behavior and the
emergent properties of the Frank model as exposed quantitatively by
applying results \cite{Janssen} from the field-theoretic
renormalization group (RG). We describe a significant result in the
field of molecular chirality, namely, symmetry breaking induced by
internal reaction noise in extremely diluted systems with absolute
enantioselective catalysis. Such a mechanism is vitally important
for current scenarios of prebiotic chemistry, where it is commonly
accepted that sufficiently high concentrations of organic compounds
could not have been reached during the chemical evolution of the
early earth.

We are interested in the long time and long wavelength properties as
governed by the nature of the RG fixed points and the topology of
the RG flow in the space of effective reaction rates. The
statistical field theory derived from the scheme
Eq.(\ref{autoLD2},\ref{mutualgeneral},\ref{decay}) maps identically
to an action for so-called multi-species directed percolation (MDP)
\cite{Janssen}, for the special case of two ``colors" or species.
Thanks to this correspondence, the full details of the RG analysis
already carried out for MDP can be carried over and applied to
analyze the critical chemical properties of the Frank model. In the
next Section, we present the field-theory action associated with the
above scheme, which after a suitable rescaling, leads to the
effective action that holds in the critical regime. In Section
\ref{RG} we reproduce the complete RG flow diagram for this model.
However, only a part of this flow diagram is applicable to real
chemical systems, and we discuss the consequences for chiral
symmetry breaking near criticality. In Section \ref{spatial} we
derive the Langevin equations that individually hold in the vicinity
of the saddle point and the two stable fixed points of that flow
diagram and integrate these numerically to obtain the time
dependence of the competing enantiomers for both large and small
noise amplitudes. The results are briefly summarized in Section
\ref{disc} where the significance of criticality for scenarios of
prebiotic chemistry is emphasized. The relation between criticality
and extremely diluted chemical systems is brought out in Appendix
\ref{dilute}. The modifications that must be made to the effective
action when the dimerization and antagonistic reactions are allowed
to be reversible are briefly discussed in Appendix \ref{reverse}.

\section{\label{Frank} The effective field theory action}

The correct inclusion of the effects of microscopic density
fluctuations in reaction-diffusion systems can be carried out once
the kinetic scheme is specified. With the scheme in hand, we derive
the corresponding chemical master equation, represent this process
by creation and annihilation operators on a spatial lattice
\cite{Doi}, and in the final step, upon taking the continuum limit,
we pass to a path integral representation \cite{Peliti,THVL}. From
this, an effective action $S_{eff}$ can be straightforwardly derived
which contains all the critical dynamics implied by the reaction
scheme to be studied. The mapping of related kinetic schemes to
continuum statistical path integrals is spelled out in
\cite{GHHT,HZ} where the main steps can be found. Applying this
procedure to the scheme in Eqs.(\ref{autoLD2}),(\ref{mutualgeneral})
and (\ref{decay}) yields the complete action $S$ governing the
reaction dynamics:
\begin{eqnarray}\label{actionahift2}
S &=& \int d^d\mathbf{x} \int dt \, \left\{ a^*\Big(
\partial_ta -D_0 \nabla^2 a+ k_2 ab -k_1A a \right. \nonumber \\
&+& k_6 a + k_3a^2 \Big) + {a^*}^2\big(k_3a^2 - k_1Aa \big)
\nonumber \\
&+&  b^*\Big(
\partial_tb -D_0 \nabla^2 b+ k_2 ab -k_1A b + k_6 b + k_3b^2 \Big) \nonumber \\
&+& {b^*}^2\big(k_3b^2 - k_1Ab \big) + k_2a^*b^*a b + k_4b^*ab
+ k_4a^*b^*ab \nonumber \\
&+& \left. k_5 a^*ab + k_5 a^*b^*ab \right\},
\end{eqnarray}
where $d$ is the spatial dimension, and
$a(\mathbf{x},t),a^*(\mathbf{x},t),b(\mathbf{x},t)$ and
$b^*(\mathbf{x},t)$ are continuous fields. In the absence of noise
(the mean field approximation) the fields
$a(\mathbf{x},t),b(\mathbf{x},t)$ correspond to the coarse-grained
local densities of the L and D enantiomers, respectively. With the
noise properly restored, these fields are generally
\textit{complex}--as is the noise--and do not directly represent the
physical densities. However, the spatial averages $\langle
a(\mathbf{x},t)\rangle,\langle b(\mathbf{x},t)\rangle$ are indeed
real and do correspond to the particle densities \cite{HT}. The
quantities $a^*(\mathbf{x},t),b^*(\mathbf{x},t)$ represent the
conjugate or response fields. These are intimately related to the
fluctuations inherent in the system. In fact, when the action $S$
depends quadratically on the conjugate fields, these can be
integrated out exactly from the path integral, and the noise
statistics completely and rigorously characterized \cite{HT}. The
\textit{non-critical} spatial dynamics (i.e., for $(k_1A - k_6)
>> 0$) implied by the action Eq.(\ref{actionahift2}) with its attendant
complex noise and fields was explored numerically in \cite{HZ} for
$k_4=k_5=0$ and $k_6 = 0$.

The fields are next rescaled in the action Eq.(\ref{actionahift2}),
for the purpose of determining which couplings (i.e., which
combinations of the rate constants $k_i$) are going to be
\textit{irrelevant} in the strict sense of the RG. This step is
needed in order to correctly identify the complete set of vertices
that are required to construct a field-theoretic perturbation
expansion of this action \cite{THVL}. We make use of the observation
that when $k_2 = 0$ and $k_4=k_5=0$, the action
Eq.(\ref{actionahift2}) reduces to that for two identical uncoupled
copies of the single particle Gribov process. We therefore rescale
the fields according to $a^* = \theta \psi^*$, $a = \theta^{-1}
\psi$, $b^* = \theta \phi^*$ and $b = \theta^{-1} \phi$, where
$\theta =(\frac{k_3}{k_1A})^{1/2}$. The space and time dependent
densities of L and D are given by $\psi$ and $\phi$, respectively.
We define the new coupling $u_0 = (k_1Ak_3)^{1/2}$. Introducing a
length scale $\kappa^{-1}$ and measuring time in units of
$\kappa^{-2}$ (i.e., $[D_0] = \kappa^0$), we find that the new
fields have scaling dimension $\kappa^{d/2}$, while $[r] =
[(k_6-k_1A)/D_0] = \kappa^2$ is a relevant perturbation in the RG
sense. On the other hand, $[u_0] = \kappa^{2-d/2}$, so this
nonlinearity becomes marginal in $d_c = 4$ dimensions. Thus, we
learn that $d=4$ is the \emph{upper critical dimension} of the Frank
model, below which the mean field approximation is incorrect. Note
that $[k_2] = [k_3] = [k_4] = [k_5] = \kappa^{2-d}$, and hence these
couplings are irrelevant compared to $[u_0]$: indeed, since e.g.,
$[k_2/u_0] = \kappa^{-d/2}$, these particular rate constants may be
omitted from the \textit{effective} action. Doing so, leads to the
effective action, which takes the form:
\begin{eqnarray}\label{effective2}
S_{eff} &=& \int d^d\mathbf{x} \int dt \, \left\{ \psi^*[\partial_t
+ D_0(r-\nabla^2)]\psi \right. \nonumber \\
&-& u_0({\psi^*}^2\psi - \psi^*\psi^2)\nonumber \\
&+& \phi^*[\partial_t + D_0(r-\nabla^2)]\phi -
u_0({\phi^*}^2\phi - \phi^*\phi^2) \nonumber \\
&+& \left. (k_2 + k_4)\theta^{-1}\phi^*\psi\phi + (k_2 +
k_5)\theta^{-1}\psi^*\psi\phi
\right\}. \nonumber \\
\end{eqnarray}
Note that the decoupling of the two enantiomers occurs when $k_2 =
k_4 = k_5 = 0$, in other words, for vanishing mutual inhibition and
dimerization. In particular, we see that both the original Frank
model and the extension treated in this paper do indeed lead to the
\textit{same} field-dependent structure for the effective action,
Eq(\ref{effective2}).

\section{\label{RG} Critical behavior}

The field theoretic renormalization group (RG) can be applied to
$S_{eff}$ in order to study the nonequilibrium critical dynamics of
this reaction-diffusion system (for a pedagogical review of this
methodology, see \cite{THVL}). The main purpose for employing RG
techniques is that they lead to differential equations describing
how the model parameters, in this case, the kinetic constants,
transform under a change of length scale. As we are here interested
in the infrared, or long wavelength, properties of the Frank model,
we therefore consider the RG flow of the parameters in the long
wavelength limit. In general, certain combinations of the kinetic
constants will flow to various fixed point (FP) values that depend
on the space dimension. Thus, the flow diagram can be constructed
revealing the critical properties of the underlying kinetic scheme.
As it turns out, $S_{eff}$ maps exactly to a field theory of
so-called multi-species directed percolation (MDP), for the special
case of two "colors" or species \cite{Janssen}. A complete and
exhaustive RG analysis has already been carried out for the general
model in \cite{Janssen}, and as pointed out there, the required
renormalization factors for MDP are provided by the single species
Gribov or directed percolation process. As an immediate consequence,
the parameter combination $(k_1Ak_3)^{1/2}/D_0 = u_0/D_0$ flows
under renormalization to the stable fixed point $u^* =
\frac{1}{2}\sqrt{2\epsilon/3}$, where $\epsilon = 4-d > 0$.

\begin{figure}[h]
\begin{center}
\includegraphics[width=0.46 \textwidth]{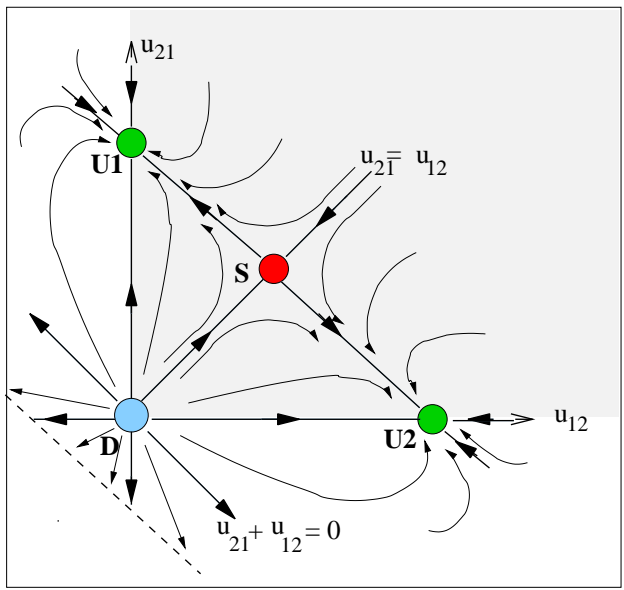}
\caption{\label{RGpoints} (Color online) RG flow or ``phase diagram"
of the reaction scheme. Space of scale dependent renormalized
couplings $u_{12}$ and $u_{21}$. The shaded quadrant corresponds to
the critical properties of the model with positive kinetic constants
and diffusion. There is an unstable fixed point \textbf{D} at
$(u_{12},u_{21}) = (0,0)$ (blue dot); a saddle point \textbf{S} at
$(u_{12},u_{21}) = (u^*,u^*)$ (red dot), and two stable fixed points
\textbf{U1} and \textbf{U2}, located at $(u_{12},u_{21}) = (0,2u^*)$
and $(u_{12},u_{21}) = (2u^*,0)$, respectively (green dots). The
separatrix $u_{21} = u_{12}$ is the boundary between the basins of
attraction of \textbf{U1} and \textbf{U2}. Outside this quadrant,
the line $u_{21} + u_{12} = 0$ separates the complete basin of
attraction of \textbf{U1} and \textbf{U2} from the dashed line of
fixed points $u_{12} + u_{21} = -2u^*$. See text and reference
\cite{Janssen} for further details.}
\end{center}
\end{figure}

We next turn to the two interspecies couplings, which from
Eq.(\ref{effective2}) are each seen to be proportional to the sum of
the rates $u_{12} \propto (k_2 + k_5)$ and $u_{21} \propto (k_2 +
k_4)$, respectively. The competition between the two enantiomers
comes in through the dependence on the rate of dimerization $k_2$,
as well as through $k_4$ and $k_5$. The complete RG analysis in
\cite{Janssen} as applied to our model proves that, except for the
point \textbf{D}, the interspecies parameters in $S_{eff}$ will flow
to one of the following $d$-dependent fixed point values:
\begin{eqnarray}\label{MDPfp} (u_{12}, u_{21}) &\equiv&
\Big(\frac{(k_2+k_5)\theta^{-1}}{D_0},
\frac{(k_2+k_4)\theta^{-1}}{D_0}\Big) \nonumber \\
&\longrightarrow& \left\{
\begin{array}{cc}
                \textbf{D}: (0,0),& \,
                \textbf{S}: (u^*,u^*) \\
                \textbf{U1}: (0,2u^*), &
                \, \textbf{U2}: (2u^*,0)
              \end{array}.
              \right.
\end{eqnarray}
The point \textbf{D} corresponds to complete decoupling between the
two enantiomers, \textbf{S} to a chiral symmetric coupling, whereas
\textbf{U1} and \textbf{U2} each correspond to homochiral final
states. The flow of the interspecies couplings $u_{12}$ and $u_{21}$
under renormalization is depicted in the flow diagram in Figure
\ref{RGpoints}. The flow, as indicated there by the sense of the
arrows, corresponds to the critical large wavelength and long time
properties of the microscopic model defined in Eqs.(\ref{autoLD2}),
(\ref{mutualgeneral}) and (\ref{decay}), and is reached for small
values of $r \approx 0$. Thus the system goes critical when the
difference in the rates of autocatalytic amplification $(k_1A)$ and
spontaneous decay $(k_6)$ goes to zero. This is achieved by varying
the concentration of the achiral matter A. This corresponds exactly
to a situation of \textit{extremely dilute} net chiral material
characterized by $\psi + \phi \approx 0$ (see Appendix \ref{dilute}
for a simple proof of this fact). Note the topology of the flow and
the stability property of each of the fixed points
(\textbf{D},\textbf{S},\textbf{U1},\textbf{U2}): (totally unstable,
saddle point, stable, and stable), respectively.

We first consider the flow properties as depicted within the
positive shaded quadrant $u_{12} > 0 $ and $u_{21} > 0$ in Fig.
\ref{RGpoints}. In so far as it is reasonable to assume that the
kinetic constants $k_i$ and the diffusion $D_0$ are nonnegative
parameters, this the most pertinent part of the full flow diagram
for real chemical systems \cite{footnote2}. There is a totally
unstable fixed point \textbf{D} located at the origin $u^*_{12} =
u^*_{21} = 0$. This corresponds to complete decoupling of the two
enantiomers. But the only way to arrive at this decoupled state is
by blithely setting the initial values of $k_2 = 0$ as well as $k_4
= k_5 = 0$ all to zero. Otherwise, the slightest positive deviation
of any one of these rates from zero, drives the system eventually to
either the saddle point \textbf{S}, if $\delta u_{12} = \delta
u_{21} > 0$, or to one of the two stable fixed points \textbf{U1},
if $\delta u_{21}
> \delta u_{12} > 0$ or \textbf{U2}, otherwise.

The diagonal symmetry line $u_{21} = u_{12}$ in this quadrant is a
separatrix dividing the basins of attraction of the two
unidirectional fixed points. For chirally symmetric kinetics, the
natural choice of course is $k_4 = k_5$, which puts the system
dynamics directly on top of this separatrix. Then, as the diagram
indicates, any positive initial value for $u_{21} = u_{12} > 0$
drives the system to the chiral symmetric fixed point \textbf{S}. In
this case, the final state of the system is determined by the fully
symmetric couplings between the two enantiomers (see
Eq.(\ref{MDPfp})). In chemistry, a \textit{racemic} mixture is one
that contains equal amounts of left- and right-handed enantiomers of
a chiral molecule. When the system is near the point \textbf{S},
racemic initial conditions lead to a racemic final state, while non
racemic initial conditions lead to a final state that maintains the
original enantiomeric excess only for low noise amplitudes (see the
numerical results in Section \ref{spatial}).

On the other hand, if the the model has $k_4 - k_5 \neq 0$ which
from Eq.(\ref{MDPfp}) implies that $u_{21} \neq u_{12}$, then the
system evolves to one of the two stable fixed points \textbf{U1} or
\textbf{U2}. At either \textbf{U1} or \textbf{U2}, the system
attains unidirectional interspecies couplings, see Eq.(\ref{MDPfp}),
which lead to the absolute amplification of one enantiomer at the
expense of the other: that is, complete chiral symmetry breaking and
a pure homochiral stable final state is the inevitable outcome (see
the numerical results in Section \ref{spatial}).

For the sake of completeness, we now address the remainder of the
flow diagram (the unshaded regions). In this case, there is then
another separatrix whose equation is $u_{12} + u_{21} = 0$, which
divides the complete basin of attraction of the two unidirectional
fixed points from the dashed line $u_{12} + u_{21} = -2u^*$; see
Figure \ref{RGpoints}. The parameter domain to the left of this
dashed line corresponds to a region of instability, and it is
conjectured in \cite{Janssen} that couplings satisfying the
condition $u_{12} + u_{21} < 0$ will lead to first order
transitions. We hasten to point out, however, that this part of the
diagram is only accessible if initial values of the $u_{12}$ and or
$u_{21}$ are negative, corresponding to a \textit{negative}
diffusion $D_0 < 0$, provided, of course, that none of the reaction
rates $k_i$ are allowed to become negative \cite{footnote2}. Thus,
the region of this diagram applicable to real chemical systems is
represented by the shaded quadrant.

\section{\label{spatial} Critical dynamics}

The temporal evolution of the two enantiomers in the critical regime
represented in Fig.\ref{RGpoints} is governed by a pair of coupled
Langevin equations which follow straightforwardly from the effective
action $S_{eff}$. These are obtained by carrying out a Gaussian
integration over the conjugate fields $\psi^*$  and $\phi^*$ in the
path integral of the exponentiated effective action: $\int
\mathcal{D}\psi \mathcal{D}{\psi^*}\mathcal{D}\phi
\mathcal{D}{\phi^*} \, e^{-S_{eff}[\psi,\psi^*,\phi,\phi^*]}$. This
final step yields a product of delta functional constraints under
the integral which in turn, lead to a pair of \textit{exact} coupled
stochastic partial differential equations \cite{HT}. The advantage
of obtaining the Langevin equations in this way is that the noise
properties are fully determined and do not have to be guessed at or
put in by hand. Numerical solutions of these stochastic equations
can be carried out to reveal the nature and qualitative tendency of
the spatial and temporal evolution of the competing enantiomers in
the neighborhood of each RG fixed point, as well as within their
respective basins of attraction.

The Langevin equations that follow from $S_{eff}$ are given by
\begin{eqnarray}\label{generalpsi}
\frac{\partial}{\partial t} \psi = D_0\nabla^2 \psi &+& ({k_1A} -
k_6)\psi -{u_0}\psi^2 \nonumber \\
&-&(k_2 + k_5)\theta^{-1}  \psi\phi
+ \xi_1 \\
\label{generalphi} \frac{\partial}{\partial t} \phi = D_0\nabla^2
\phi &+& ({k_1A}-k_6)\phi - {u_0}\phi^2 \nonumber \\
&-& (k_2 +
k_4)\theta^{-1} \phi\psi + \xi_2,
\end{eqnarray}
where the noise satisfies $\langle \xi_1 \rangle = \langle \xi_2
\rangle = 0$ and
\begin{eqnarray}\label{gennoisepsi}
\langle \xi_1(\mathbf{x},t) \xi_1(\mathbf{x'}, t')\rangle &=& 2
{u_0}\psi(\mathbf{x}, t)\delta^{d}(\mathbf{x}-\mathbf{x}')\delta(t
-t'), \\ \label{gennoisephi} \langle \xi_2(\mathbf{x}, t)
\xi_2(\mathbf{x'}, t')\rangle &=& 2 {u_0}\phi(\mathbf{x},
t)\delta^{d}(\mathbf{x}-\mathbf{x}')\delta( t -  t').
\end{eqnarray}
These equations hold in the critical region shown in Fig.
\ref{RGpoints}.

\subsection{\label{Spoint} Langevin equations in the vicinity of the saddle point}

In particular, the behavior of the model in the vicinity of the
saddle point \textbf{S} where the interspecies couplings $u_{12}$
and $u_{21}$ flow to a symmetric fixed value, is given by the
solutions of the system Eqs.(\ref{generalpsi},\ref{generalphi})
where we now set $k_4 = k_5$. These are subject to the noise
Eqs.(\ref{gennoisepsi},\ref{gennoisephi}), and we use the result
that $u_0/D_0$ flows to $u^*$, together with the corresponding fixed
point values for $u_{12}$ and $u_{21}$, as given in
Eq.(\ref{MDPfp}). At this juncture, it is also convenient to rescale
the fields $\tilde \psi = D_0u^*/(k_1A-k_6)\psi$, $\tilde \phi =
D_0u^*/(k_1A-k_6)\phi$, and employ dimensionless time $\tau =
(k_1A-k_6) t$ and coordinates $\hat x_j = ((k_1A-k_6)/D_0)^{1/2}
x_j$. These simple steps yield the stochastic equations in the
vicinity of the saddle point \textbf{S}:
\begin{eqnarray}\label{Sdynamics1}
\partial_{\tau} \tilde \psi &=& \hat \nabla^2 \tilde \psi + \tilde \psi - \tilde \psi^2 -\tilde \psi
\tilde \phi + \tilde \xi_1(\mathbf{\hat x}, \tau), \\
\label{Sdynamics2}
\partial_{\tau} \tilde \phi &=& \hat \nabla^2 \tilde \phi + \tilde \phi - \tilde \phi^2 -\tilde \phi
\tilde \psi + \tilde \xi_2(\mathbf{\hat x}, \tau),
\end{eqnarray}
where the rescaled noise is given by
\begin{eqnarray}\label{Snoisepsi}
\langle \tilde \xi_1(\mathbf{\hat x},\tau) \tilde \xi_1(\mathbf{\hat
x}', \tau')\rangle &=& 2\Big(\frac{D_0}{k_1A -
k_6}\Big)^{2-d/2}{u^*}^2 \tilde \psi(\mathbf{\hat x}, \tau)\nonumber
\\ &\times&
\delta^{d}(\mathbf{\hat x}-\mathbf{\hat x}')\delta(\tau -\tau'), \\
\label{Snoisephi} \langle \tilde \xi_2(\mathbf{\hat x},\tau) \tilde
\xi_2(\mathbf{\hat x}', \tau')\rangle &=& 2\Big(\frac{D_0}{k_1A -
k_6}\Big)^{2-d/2}{u^*}^2\tilde \phi(\mathbf{\hat x}, \tau)\nonumber
\\ &\times& \delta^{d}(\mathbf{\hat x}-\mathbf{\hat x}')\delta(\tau -\tau').
\end{eqnarray}
In two dimensions, the noise strength is characterized by the
parameter $\sigma^2 = 2D_0 {u^*}^2/(k_1A - k_6)$, with $u^* =
1/\sqrt{3} \approx 0.58$, and this can be large or small depending
on whether the diffusion rate $D_0$ is large or small (keeping the
difference $k_1A-k_6 > 0$ fixed), respectively. In Figure
\ref{Spoint}, some representative effects of the diffusion on the
critical dynamics in the neighborhood of the saddle point \textbf{S}
in $d=2$ are displayed for the spatially averaged enantiomers for
both large $(\sigma = 1)$ and small $(\sigma = 0.3)$ internal noises
and for non-racemic initial conditions. Recall that here we have set
$k_4 = k_5$. We solve numerically the full stochastic
two-dimensional version of Eqs. (\ref{Sdynamics1},\ref{Sdynamics2})
subject to the noise given by Eqs. (\ref{Snoisepsi},\ref{Snoisephi})
, using reflecting boundary conditions and a finite difference
scheme with $\Delta \tau=0.005$, $\Delta \hat{x}=\Delta
\hat{y}=0.23$, and a grid of size $L \times L =154\times154$.

\begin{figure}[h]
\begin{center}
\includegraphics[width=0.5 \textwidth]{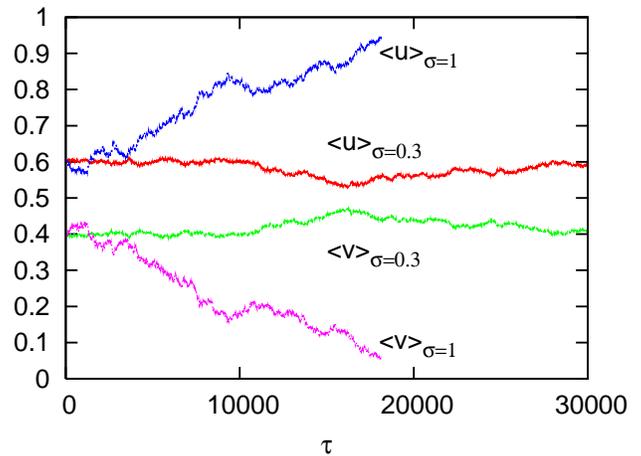}
\caption{\label{Spoint} (Color online) $k_4 = k_5$: Time evolution
of the spatially averaged enantiomer densities $\langle u \rangle =
\langle \tilde \phi \rangle$ and of $\langle v \rangle = \langle
\tilde \psi \rangle$ in two dimensions $d=2$ for two representative
simulations of the stochastic dynamics near the saddle point
\textbf{S}, Eqs.(\ref{Sdynamics1},\ref{Sdynamics2}) with noise
Eqs.(\ref{Snoisepsi},\ref{Snoisephi}). Evolution of non-racemic
homogeneous initial conditions $\big((\tilde \phi(\hat x,\hat y,\tau
= 0),\tilde \psi(\hat x, \hat y,\tau = 0)\big) = (0.6,0.4)$. For
weak noise $\sigma = 0.3$ see the inner pair of red (dark gray) and
green (light gray) curves. For stronger $\sigma = 1$ noise, see the
outer pair of blue (dark gray) and magenta (light gray) curves.}
\end{center}
\end{figure}
For small noise levels $(\sigma = 0.3)$ , and for non-racemic
initial compositions, the initial proportion of the two chiral
species is roughly maintained, modulo the fluctuations; see the
inner pair of curves in Fig. \ref{Spoint}. The evolution of $\langle
u \rangle_{\sigma=0.3}$ is shown in the red (dark gray) line  and
$\langle v \rangle_{\sigma=0.3}$ in the green (light gray) line.
However, for stronger noise $(\sigma = 1)$, the initial imbalance
shows an almost monotonic tendency to increase, suggesting that
sufficiently strong noise is capable of driving the system to a
homochiral final state, in spite of the manifest mathematical chiral
symmetry of the underlying evolution equations and noise terms under
the substitutions $\tilde \psi \rightarrow \tilde \phi$ and $\tilde
\phi \rightarrow \tilde \psi$. A mean field analysis of the
solutions of Eqs.(\ref{Sdynamics1}) and (\ref{Sdynamics2}), which
ignores both diffusion and noise, indicates that the enantiomeric
excess of the concentrations of the two enantiomers is time
independent \cite{GTV}. Here, this is seen to be approximately true
also for the spatially averaged diffusing enantiomers subject to
small noise. But greater internal noise induces a striking departure
from this that is not captured by the mean field approximation, as
seen in Fig. \ref{Spoint}. This is depicted in the outer pair of
curves. The evolution of $\langle u \rangle_{\sigma=1}$ is shown in
the blue (dark gray) line  and $\langle v \rangle_{\sigma=1}$ in the
magenta (light gray) line.

As we are interested here in displaying only the initial and
intermediate time dependent tendencies of the two enantiomer
densities in the vicinities of the various RG fixed points, we have
employed standard integration of the Langevin equations, sufficient
for revealing the qualitative nature of the solutions for short and
intermediate computational time steps, as can be seen in Fig. 2 and
Fig. 3 (below). Near an absorbing state transition, one of the
densities tends to zero, and the numerical integration breaks down.
This can be seen clearly in the simulation of the evolution to the
absorbing state, which is shown only for the shorter time scales in
these figures.  This standard algorithm is of course not adequate
for extracting the much more precise and delicate information such
as asymptotic decays or power law exponents.  For the latter, we
would have had to appeal to the more sophisticated numerical schemes
such as those proposed by Dickman \cite{Dickman}, Moro \cite{Moro}
or by Dornic et al. \cite{Dornic}.

\subsection{\label{Upoints} Langevin equations in the vicinity of the unidirectional fixed points}
\begin{figure}[h]
\begin{center}
\includegraphics[width=0.5 \textwidth]{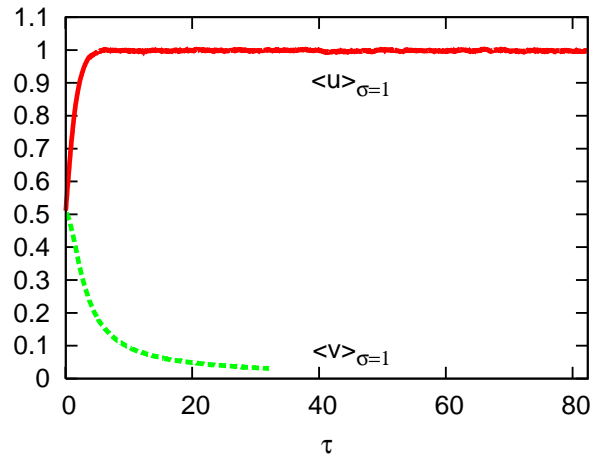}
\caption{\label{homo1} (Color online) $k_4 \neq k_5$: Time evolution
of the spatially averaged enantiomer densities $\langle u \rangle =
\langle \tilde \phi \rangle$ and of $\langle v \rangle = \langle
\tilde \psi \rangle$ in two dimensions $d=2$ for a representative
simulation of the stochastic dynamics near the asymmetric
unidirectional fixed point \textbf{U1},
Eqs.(\ref{Udynamics1},\ref{Udynamics2}). Evolution of homogeneous
racemic initial conditions $\big((\tilde \phi(\hat x,\hat y,\tau =
0),\tilde \psi(\hat x, \hat y,\tau = 0)\big) = (0.5,0.5)$ for noise
level $\sigma = 1$. $\langle u \rangle_{\sigma = 1}$ is shown in the
upper solid red (dark gray) curve and $\langle v \rangle_{\sigma =
1}$ is shown in the lower dashed green (light gray) curve.}
\end{center}
\end{figure}

The behavior of the system near one of the two stable attracting
fixed points, for instance \textbf{U1}, is determined by the pair of
equations
\begin{eqnarray}\label{Udynamics1}
\partial_{\tau} \tilde \psi &=& \hat \nabla^2 \tilde \psi
+ \tilde \psi - \tilde \psi^2 - 2 \tilde \psi
\tilde \phi + \tilde \xi_1(\mathbf{\hat x}, \tau), \\
\label{Udynamics2}
\partial_{\tau} \tilde \phi &=& \hat \nabla^2 \tilde \phi
+ \tilde \phi - \tilde \phi^2 + \tilde \xi_2(\mathbf{\hat x}, \tau),
\end{eqnarray}
with the noise properties as given above in
Eqs.(\ref{Snoisepsi},\ref{Snoisephi}). Here, we use the fixed point
value $(u_{12},u_{21}) = (0,2u^*)$. Recall in order to arrive at
this fixed point, we set $k_4 \neq k_5$. This corresponds to
``starting" the system off in either the basin of attraction of
\textbf{U1} or that of \textbf{U2} (see shaded quadrant in Fig.
\ref{RGpoints}). Note the manifest asymmetry in the equation pair
due to the presence of the unidirectional coupling term in
Eq.(\ref{Udynamics1}), absent from Eq.(\ref{Udynamics2}). This fixed
point is associated with homochirality, as confirmed by numerical
simulation; see Fig. \ref{homo1}. Starting from racemic initial
conditions, the plot of the spatially averaged enantiomeric
densities, in Fig. \ref{homo1}, indicates an extremely rapid onset
of absolute chiral amplification.

The evolution of $\langle u \rangle_{\sigma = 1}$ is shown in the
upper solid red (dark gray) curve and that of $\langle v
\rangle_{\sigma = 1}$ is shown in the lower dashed green (light
gray) curve. The equations governing the critical dynamics at the
other stable fixed point \textbf{U2} are had by simply interchanging
the fields $\tilde \psi \leftrightarrow \tilde \phi$ in
Eqs(\ref{Udynamics1},\ref{Udynamics2}) and replacing $\tilde \xi_1
\leftrightarrow \tilde \xi_2$. The behavior is quantitatively
identical, but with the roles of the two enantiomers obviously
reversed.

Near criticality, there is no numerical evidence for the formation
of the spatially segregrated chiral domains bounded by racemic
fronts, in marked contrast to the results reported in \cite{HZ}. The
distinguishing factor of course is that the present simulations are
carried out at the unidirectional critical points \textbf{U1} or
\textbf{U2}, whereas in \cite{HZ}, the system was explored
\textit{far away} from criticality, where such chirally pure domains
are expected to form \cite{Frank,Decker}. At the RG unidirectional
fixed points, the dynamical equations themselves are manifestly
chirally asymmetric, and the system rapidly evolves to a final
homochiral state, without passing through the intermediate stages of
enantiomeric competition.

-----------------------------------------------------------
\subsection{\label{Upoints} Stochastic dynamics away from the critical points}

The distinction between the critical and non-critical behavior can
be sharpened by contrasting the mathematical structure of the
critical equations and noise to those that hold \textit{away} from
the fixed points. The latter are of course given by the system
Eqs.(\ref{generalpsi},\ref{generalphi}) with the fluctuations
obeying Eqs.(\ref{gennoisepsi},\ref{gennoisephi}) \cite{footnote}.
By means of the field rescaling  $\tilde \psi = ((k_2 + k_5)/u_0)
\psi$, and with a similar relation between $\tilde \phi$ and $\phi$,
these Langevin equations can be written as (note: in what follows we
take $k_4=k_5$)
\begin{eqnarray}\label{NCdynamics1}
\partial_{\tau} \tilde \psi &=& \hat \nabla^2 \tilde \psi
+ \tilde \psi - g \tilde \psi^2 -\tilde \psi
\tilde \phi + \eta_1(\mathbf{\hat x}, \tau), \\
\label{NCdynamics2}
\partial_{\tau} \tilde \phi &=& \hat \nabla^2 \tilde \phi
+ \tilde \phi - g \tilde \phi^2 -\tilde \phi
\tilde \psi + \eta_2(\mathbf{\hat x}, \tau),
\end{eqnarray}
where $g = k_3/(k_2 + k_5)$ and the rescaled noise obeys

\begin{eqnarray}\label{NCnoisepsi}
\langle  \eta_1(\mathbf{\hat x},\tau) \eta_1(\mathbf{\hat x}',
\tau')\rangle &=& 2\frac{(k_2+ k_5)}{D_0^{d/2}}(k_1A-k_6)^{d/2-1}
\tilde \psi(\mathbf{\hat x}, \tau)\nonumber \\
&\times&
\delta^{d}(\mathbf{\hat x}- \mathbf{\hat x}')\delta(\tau -\tau'), \\
\label{NCnoisephi} \langle \eta_2(\mathbf{\hat x},\tau)
\eta_2(\mathbf{\hat x}', \tau')\rangle &=& 2 \frac{(k_2+
k_5)}{D_0^{d/2}}(k_1A-k_6)^{d/2-1}\tilde \phi( \mathbf{\hat x},
\tau)\nonumber \\
&\times& \delta^{d}(\mathbf{\hat x}- \mathbf{\hat x}')\delta(\tau
-\tau').
\end{eqnarray}

If the noise and the diffusion terms are ignored, then a mean field
analysis of the homogeneous asymptotic solutions of the
corresponding equations Eqs.(\ref{NCdynamics1},\ref{NCdynamics2}),
reveals that the parameter $g$ plays a special role \cite{GTV}.
Indeed, $g<1$ leads to chiral amplification of an initial
enantiomeric excess, whereas $g>1$ leads to a racemic final state.
The point $g=1$ was identified as a \textit{critical} value, in the
sense that a sudden qualitative change in the asymptotic behavior of
the mean-field solutions is observed: the ratio of the two
enantiomeric concentrations in this borderline case remains
constant, and depends on the initial composition. If the initial
condition is racemic, the system will always remain racemic, if
however there is a slight initial excess, this excess is forever
maintained \cite{GTV}.

From these remarks we see that the spatially dependent and
stochastic effective equations associated with each renormalization
group fixed point $\mathbf{S,U1}$,
Eqs.(\ref{Sdynamics1},\ref{Sdynamics2}) and
Eqs.(\ref{Udynamics1},\ref{Udynamics2}), respectively, have $g=1$.
Likewise for the point \textbf{U2}. It is \textit{as if} we had set
$g$ to its ``critical" value. But it is important to emphasize that
at these RG fixed points, $g$ is no longer a freely adjustable
parameter, but under renormalization is automatically driven to this
special value. The RG thus provides a rational physical explanation
for why $g=1$ at criticality.

\section{\label{disc} Discussion}

As supported by surveys and reviews of the present status of chiral
autocatalysis, mirror symmetry breaking, stochasticity, and their
implications for the origin of homochirality, the Frank model and
its extensions continue to serve as the paradigm for theoretical
studies of this phenomenon \cite{Kondepudi,Blackmond,WLL,Ribo}. In
this paper we have studied the critical properties of the Frank
model and a simple extension of it, by exploiting the mapping of
this kinetic scheme to a well studied phenomenon from condensed
matter and non-equilibrium statistical physics, namely,
(multi-species) directed percolation processes \cite{Janssen}. By
virtue of this exact mapping, which is established at the level of
statistical field theory, the complete renormalization group (RG)
analysis of the critical properties of directed percolation can be
applied to study the critical features of the Frank model. The most
significant result in this paper is the  description of a new effect
in the field of molecular chirality, namely mirror symmetry breaking
induced by internal noise in extremely diluted systems with absolute
enantioselective catalysis. Such a mechanism is of course vitally
important for scenarios of prebiotic chemistry, where it is commonly
agreed upon that sufficiently high concentrations of organic
compounds could not have been reached during the chemical evolution
of the early earth \cite{prebio}. These final states are consequence
of internal composition fluctuations and reactions limited by
spatial diffusion. To reach these dilute multi-critical states, the
difference between the amplification and decay rates must be close
to zero \cite{Janssen}. This contrasts to Saito and Hyuga's
suggestion that, for closed systems,  both \textit{nonlinear}
autocatalysis and recycling with diffusion seem to be required for
chiral symmetry breaking in dilute solutions \cite{SH}. Additional
insight into the dynamical consequence of each transition is
provided by deriving and numerically solving the exact
\textit{effective} Langevin equations that hold in the neighborhood
of each fixed point.

It has been known for some time that the chirally symmetric state of
the Frank model is unstable and that (external) fluctuations can
induce a transition to homochiral final states. There are evidently
a number of distinct routes leading to homochirality, and the
concept of criticality and the identification of the associated
critical parameters should be clearly distinguished. From general
bifurcation theory, we thus learn that the transition from a
symmetric to a chiral final state can be induced by varying solely
the concentrations of the substrate molecules \cite{KN}. The mean
field analysis of Ref. \cite{GTV} on the other hand, identifies the
ratio of rate constants $ g = k_3/k_2$ as the pertinent critical
parameter. In certain crystallization experiments, it is the
\textit{stirring rate} of the solution that has been observed to
play the role of a critical parameter \cite{Ketal}. The present work
makes use of the fact that chiral or mirror symmetry breaking is an
example of an active to absorbing state phase transition, and that
such transitions are generically characterized by directed
percolation processes (DP)\cite{THVL}. RG techniques can be applied
to analyze this symmetry breaking phenomenon in extremely dilute
chemical systems in a precise manner.

\begin{acknowledgments}
We acknowledge useful conversations with Prof. Josep M. Rib\'{o}
during the course of this work. We are especially indebted to Prof.
Vladik Avetisov for recognizing the correct significance of our
results in the context of prebiotic chemistry. We thank Dr. Carlos
Briones for supplying us with second book cited in Ref
\cite{prebio}. This research is supported in part by the Grant
AYA2006-15648-C02-02 from the Ministerio de Educaci\'{o}n y Ciencia
(Spain).
\end{acknowledgments}

\appendix
\section{\label{dilute}Critical regime implies extreme dilution}

Simple bifurcation analysis of the purely kinetic scheme
Eqs.(\ref{autoLD2},\ref{mutualgeneral},\ref{decay})
\begin{eqnarray}\label{Lone}
\frac{dL}{dt} &=& (k_1A - k_6)L - k_3L^2 - (k_2 + k)LD,\\
\label{Done} \frac{dD}{dt} &=& (k_1A - k_6)D - k_3D^2 - (k_2 + k)LD,
\end{eqnarray}
reveals the chemical nature of the critical regime of the fully
stochastic field theory treated in this paper \cite{thanks}. Here,
$A, L$ and $D$ denote concentrations and we have set $k_4 = k_5 =
k$. Introduce the enantiomeric excess $\eta = \frac{L - D}{L + D}$
and the total concentration of chiral matter $\chi = L + D$. Then
the kinetic equations Eqs.(\ref{Lone},\ref{Done}) can be written as
follows:
\begin{eqnarray}\label{eta}
\frac{d\eta}{dt} &=& \frac{k_2 + k - k_3}{2}\chi\eta(1 - \eta^2)\\
\label{chi} \frac{d \chi}{dt} &=&  (k_1A-k_6)\chi - [ k_3 +
\frac{k_2 +
k - k_3}{2}(1 - \eta^2)]\chi^2. \nonumber \\
\end{eqnarray}
For behavior with large amount of chiral matter, $\chi >> 0$, we
must have $(k_1A - k_6) >> 0$. This requires that the system be
\textit{far} from criticality. The bifurcation equation \cite{AG} is
then $\eta(1- \eta^2) = 0$ and there are three stationary solutions:
\begin{equation}
\{ \eta = 0, \chi = 2\frac{k_1A - k_6}{k_2 + k + k_3} \},
\end{equation}
unstable if $k_2 - k >0$ and stable if $k_2- k<0$, and
\begin{equation}
\{ \eta = \pm 1, \chi = \frac{k_1A - k_6}{k_3} \},
\end{equation}
stable if $k_2 - k > 0$ and unstable if $k_2 - k < 0$. From this we
can deduce the following salient features: First, by introducing the
dimensionless time $\tau = (k_1A - k_6) t$, we see that far from
criticality, the variable $\chi$ changes much more rapidly than the
enantiomeric excess $\eta$. The system rapidly reaches a
quasistationary state for $\chi$ $(d\chi/d\tau \approx 0)$ and then
the slow variable $\eta$ evolves and the full system reaches its
true steady state. Secondly, the criticality condition $(k_1A - k_6)
\approx 0$ corresponds to the kinetic behavior under conditions of
extreme dilution, $\chi \approx 0$, with the concentration of chiral
material close to zero. In this case, the system has no well defined
steady state with respect to the concentration $\chi$, since $\eta =
(L-D) /\chi$ yields an indeterminate expression near criticality.
Furthermore, the equation for $\chi$ becomes as ``slow" as the
equation for $\eta$, and the use of classical kinetic approach based
on the law of mass action becomes questionable. Thus, the stochastic
approach employed in this paper is not only justified, but is needed
to correctly describe the critical regime of the Frank model.

\section{\label{reverse}Reversible reactions}

The backreaction of the dimerization step is eliminated if the
product $P$ is continuously being removed from the reactor.
Otherwise, the reverse reaction must be taken into account.
Furthermore, as pointed out by Avetisov and Goldanskii \cite{AG},
reversibility in the mutual inhibition reactions will account for
the limited enantioselectivity of chiral catalysts, so that the
catalytic effect of each enantiomer leads to the formation of
\textit{both} L and D products. To include these effects, the
following reactions would have to be added to the above scheme
Eqs.(\ref{autoLD2},\ref{mutualgeneral},\ref{decay}):
\begin{equation}\label{reversableP}
P \stackrel{k_{-2}}{\longrightarrow} \textrm{L} + \textrm{D}, \qquad
\textrm{L} + \textrm{A} \stackrel{k_{-4}}{\longrightarrow}
\textrm{L} + \textrm{D}, \qquad \textrm{D} + \textrm{A}
\stackrel{k_{-5}}{\longrightarrow} \textrm{L} + \textrm{D}.
\end{equation}
The lefthandmost corresponds to the backreaction of the
dimerization, while the latter two allow for reversibility in the
mutual inhibition reactions.  Going through the same algebraic
procedure that led us to the effective action in Sec \ref{Frank}, we
find that the above reactions yield the following terms to be added
to the effective action in Eq.(\ref{effective2}):
\begin{eqnarray}\label{deltaSeff}
\Delta S_{eff} &=& -\int d^d\mathbf{x} \int dt \left\{
k_{-2}\theta\psi^* + k_{-2}\theta\phi^* \right. \nonumber \\
&+& k_{-2}\theta^2\psi^*\phi^* +  k_{-4} \phi^*\psi + k_{-5}
\psi^*\phi  \\ \nonumber &+& \left.  k_{-4}\theta\psi^*\psi\phi^*
k_{-5}\theta\phi^*\phi\psi^* \right\}.
\end{eqnarray}
From dimensional analysis we find that $[k_{-2}\theta] =
\kappa^{2+d/2}$, $[k_{-2}\theta^2] = [k_{-4}] = [k_{-5}] = \kappa^2$
are relevant perturbations in the sense of the RG for all
dimensions, whereas $[k_{-4}\theta] = [k_{-5}\theta] =
\kappa^{2-d/2}$. The corresponding cubic terms are therefore
marginal in $d_c =4$ dimensions and are relevant for $d<4$.

From the point of view of the field-theoretic content of $\Delta
S_{eff}$, we see that the dimerization backreaction induces new
relevant terms not present in the original effective action
proportional to  $\sim\psi^*, \sim\phi^*$, as well as a term of
dimension $\kappa^2$. However, this additional term proportional to
$\psi^*\phi^*$ dynamically couples the two enantiomers via a
cross-correlated noise, a feature not present in the absence of
dimer breakup. Regarding the limited enantioselectivity reactions,
these induce ``masslike" terms $\sim \phi^*\psi, \sim\psi^*\phi$,
that also serve to link the two enantiomers. In fact, these terms
lead to off diagonal contributions to the response functions or
propagators. The new cubic terms $\sim\psi^*\psi\phi^*
,\sim\phi^*\phi\psi^*$ are also sources of ``off-diagonal" or
cross-correlated reaction noise; in graphical perturbation theory,
these lead to new cubic vertices which would have to be included in
a field-theoretic RG analysis, such as in \cite{GHHT}.


\begin{thebibliography}{99}
\bibitem{AG} V. Avetisov and V. Goldanskii, Proc. Natl. Acad. Sci.
USA, \textbf{93}, 11435 (1996).
\bibitem{Avalos2} M. Avalos, R. Babiano, P. Cintas, J.L. Jim\'{e}nez and
J.C. Palacios, Tetrahedron Asymm. \textbf{11}, 2845 (2000).
\bibitem{Kondepudi} D.K. Kondepudi and K. Asakura, Acc. Chem. Res.
\textbf{34}, 946 (2001).
\bibitem{Blackmond} D.G. Blackmond, Proc. Natl. Acad. Sci. USA \textbf{101}, 5732
(2004).
\bibitem{WLL} I. Weissbuch, L. Leiserowitz and M. Lahav, Top. Curr.
Chem. \textbf{259}, 123 (2005).
\bibitem{Frank} F.C. Frank, Biochim. et Biophys. Acta \textbf{11}, 459 (1953).
\bibitem{footnote1} In this paper, the symbols ``L"  (lefthanded) and ``D" (righthanded) are used
to refer only to the molecule's geometrical or spatial conformation.
Handedness, or chirality,  is an extrinsic property. Crystal
chirality can be determined from its optical activity. However, and
this is the subtle point to be aware of, some lefthanded molecules
rotate plane polarized light in the clockwise sense, while others do
so in the anticlockwise sense (and similarly for righthanded
molecules). For this reason, we refrain from employing the
terminology ``levo" and ``dextro", which refer strictly to the
\textit{intrinsic} optical properties of (chiral) molecules.
\bibitem{Sandars} P.G.H. Sandars, Orig. Life Evol. Biosph. \textbf{33}, 575
(2003).
\bibitem{BAHN} A. Brandenburg, A.C. Andersen, S. H\"{o}fner and M.
Nilsson, Orig. Life Evol. Biosph. \textbf{35}, 225 (2005).
\bibitem{BM} A. Brandenburg and T. Multam\"{a}ki, Int. J. Astrobiol. \textbf{3},
209 (2004).
\bibitem{GT} M. Gleiser and J. Thorarinson, Orig. Life Evol. Biosph.
\textbf{36}, 501 (2006).
\bibitem{Gleiser} M. Gleiser, Orig. Life Evol. Biosph. \textbf{37}, 235
(2007).
\bibitem{Janssen} H.-K. Janssen, Jour. Stat. Phys. \textbf{103}, 801
(2001); Phys. Rev. Lett. \textbf{78}, 2890 (1997).
\bibitem{Doi} M. Doi, J. Phys. A: Math. Gen.
\textbf{9}:1465, 1479 (1976).
\bibitem{Peliti} L. Peliti, J. Physique \textbf{46}:1469 (1985).
\bibitem{THVL}
U.C. T\"{a}uber, M. Howard and B.P. Volmayr-Lee, J. Phys. A: Math.
Gen. \textbf{38}, R79 (2005).
\bibitem{GHHT} Y.Y. Goldschmidt, H. Hinrichsen, M. Howard, and U.C.
T\"{a}uber, Phys. Rev. E \textbf{59}, 6381 (1999).
\bibitem{HZ} D. Hochberg and M.-P. Zorzano, Chem. Phys. Lett.
\textbf{431}, 185 (2006).
\bibitem{HT} M.J. Howard and U.C.
T\"{a}uber, J. Math A: Gen. \textbf{30}, 7721 (1997).
\bibitem{footnote2} In chemistry, there are no negative rate
constants. As for negative diffusion, Fokker-Planck (FP) equations
for certain nonlinear chemical and optical systems can have
non-positive-definite diffusion coefficients. These FP's can however
be made sense of, and a rigorous justification provided. An early
paper reviewing the correct treatment of negative $D$ is given in
the paper by P.D. Drummond, C.W. Gardiner and D.F. Walls, Phys. Rev
A \textbf{24}, 914 (1981).
\bibitem{GTV} I. Gutman, D. Todorovi\'{c} and M. Vu\v{c}kovi\'{c}, Chem. Phys.
Lett. \textbf{216}, 447 (1993).
\bibitem{Dickman} R. Dickman, Phys. Rev. E \textbf{50}, 4404 (1994).
\bibitem{Moro} E. Moro, Phys. Rev E \textbf{70}, 045102(R) (2004).
\bibitem{Dornic} I. Dornic, H. Chat\'{e}, and M.A. Mu\~{n}oz, Phys. Rev.
Lett. \textbf{94} 100601 (2005).
\bibitem{Decker} P. Decker, in \textit{Origins of optical activity in
nature}, edited by David C. Walker (Elsevier, New York, 1979), pg
109-124.
\bibitem{footnote} Apart from the noise terms, the rescaled Langevin equations
for the Frank model that hold \textit{away} from the critical regime
are of course the same as those derived in Ref \cite{HZ}. The only
difference comes in through the noise terms. But the mean field
analysis yields the same results with the parameter $g$ playing the
role as indicated in Ref \cite{GTV}.
\bibitem{Ribo} J. Crusats, S. Veintemillas-Verdaguer and J.M. Rib\'{o},
Chem. Eur. J., \textbf{12}, 7576 (2006).
\bibitem{prebio} S.F. Mason, \textit{Chemical Evolution} (Oxford,
1991); O. Botta and J.L. Bada, in \textit{The Genetic Code and the
Origin of Life}, edited by Ll. Ribas de Pouplana (Kluwer
Academic/Plenum, New York, 2004).
\bibitem{SH} Y. Saito and H. Hyuga, J. Phys. Soc. Jpn., \textbf{73}, 1685
(2004).
\bibitem{KN} D.K. Kondepudi and G.W. Nelson, Physica \textbf{125}A, 465 (1984).
\bibitem{Ketal} D.K. Kondepudi, K. L. Bullock, J.A. Digits, and P.D.
Yarborough, J. Am. Chem. Soc. \textbf{117}, 401 (1995).
\bibitem{thanks} We are grateful to Prof. V. Avetisov for providing the
calculation summarized in this Appendix and its attendant
interpretation.
\end{thebibliography}
\end{document}